\documentclass[letterpaper,conference]{ieee/IEEEtran}

\usepackage{amssymb}
\usepackage[cmex10]{amsmath}

\usepackage[capitalize]{cleveref}

\usepackage{float}

\usepackage{graphicx}

\usepackage{subcaption}

\usepackage{calc}

\usepackage{wrapfig}

\usepackage{tabularx}
\usepackage{booktabs}
\usepackage{footnote}

\usepackage{makeidx}        

\usepackage{balance} 

\usepackage{textcomp}

\usepackage{booktabs}

\usepackage{color}

\usepackage{colortbl}
\usepackage[dvipsnames]{xcolor}

\usepackage{siunitx}

\usepackage[acronym]{glossaries}
\newglossary[symlog]{symbol}{symi}{symo}{Symbols}
\glstoctrue 
\newcommand{\newacronymf}[3]{\newglossaryentry{#1}{name={#2},description={#3},first={#2}}}

\newacronym{3gpp}{3GPP}{Third Generation Partnership Project}
\newacronym{4g}{4G}{fourth generation of mobile phone system specifications}
\newacronym{5g}{5G}{fifth generation of mobile phone system specifications}
\newacronym{1dlstm}{1D-LSTM}{One-dimensional Long Short-Term Memory}
\newacronym{2dlstm}{2D-LSTM}{Two-dimensional Long Short-Term Memory}
\newacronym{1dcnn}{1D-CNN}{One-dimensional Convolutional Neural Network}
\newacronym{2dcnn}{2D-CNN}{Two-dimensional Convolutional Neural Network}

\newacronym{ann}{ANN}{Artificial Neural Network}
\newacronym{anns}{ANNs}{Artificial Neural Networks}
\newacronym{acq}{ACQ}{acquisition}
\newacronym{app}{APP}{a posteriori probability}
\newacronym{ask}{ASK}{amplitude-shift keying}
\newacronym{awgn}{AWGN}{Additive White Gaussian Noise}
\newacronym{arq}{ARQ}{automatic repeat request}
\newacronym{acs}{ACS}{add-compare-select}
\newacronym{asic}{ASIC}{Application-Specific Integrated Circuit}
\newacronym{arp}{ARP}{almost regular permuatation}
\newacronym{asip}{ASIP}{application specific instruction set processor}
\newacronym{amba}{AMBA}{Advanced Microcontroller Bus Architecture}
\newacronym{ahb}{AHB}{advanced high-performance bus}
\newacronym{aacs}{AACS}{a priori aware constant sphere}
\newacronym{ant}{ANT}{algorithmic noise-tolerance}
\newacronym{ai}{AI}{Artificial Intelligence}
\newacronym{af}{AF}{Atrial Fibrillation}
\newacronym{aecg}{AECG}{Ambulatory Electrocardiographic}

\newacronym{bhl}{BHL}{Backward Hidden Layer}
\newacronym{bcjr}{BCJR}{Bahl, Cocke, Jelinek, and Raviv}
\newacronym{bpsk}{BPSK}{Binary Phase Shift Keying}
\newacronym{ber}{BER}{bit error rate}
\newacronym{bp}{BP}{Belief Propagation}
\newacronym{bw}{BW}{Bit Width}
\newacronym{bicm}{BICM}{bit interleaved coded modulation}
\newacronym{blast}{BLAST}{Bell Labs layered space time}
\newacronym{bc}{BC}{best case, 1.3V, -40$^\circ$C}
\newacronym{bilstm}{Bi-LSTM}{Bidirectional LSTM}
\newacronym{bmbf}{BMBF}{Federal Ministry of Education and Research}
\newacronym{bnn}{BNN}{Binarized Neural Network}
\newacronym{bram}{BRAM}{Block RAM}

\newacronym{ctc}{CTC}{Connectionist Temporal Classification}
\newacronym{cn}{CN}{Check Node}
\newacronym{cr}{CR}{Code Rate}
\newacronym{cfu}{CFU}{Check Node Functional Unit}
\newacronym{crc}{CRC}{Cyclic Redundancy Check}
\newacronymf{cdma2000}{CDMA2000}{name of a communication standard}
\newacronym{cc}{CC}{convolutional code}
\newacronym{cnb}{CNB}{check node block}
\newacronym{cnp}{CNP}{check node processor}
\newacronym{cdma}{CDMA}{code division multiple access}
\newacronym{cmos}{CMOS}{complementary metal oxide semiconductor}
\newacronym{cpu}{CPU}{Central Processing Unit}
\newacronym{cnn}{CNN}{Convolutional Neural Network}
\newacronym{cer}{CER}{Character Error Rate}
\newacronym{ccl}{CCL}{Connected Component Labeling}
\newacronym{clb}{CLB}{Configurable Logic Block}

\newacronym{darts}{DARTS}{Differentiable ARchiTecture Search}
\newacronym{dart}{DART}{Distributed Analytics Runtime}
\newacronym{dma}{DMA}{Direct Memory Access}
\newacronym{dnnu}{DNNU}{Deep Neural Network Unit}
\newacronym{db}{dB}{decibel}
\newacronym{dp}{DP}{dual ported (memory)}
\newacronym{dram}{DRAM}{Dynamic Random Access Memory}
\newacronym{dc}{$d_c$}{CN degree}
\newacronym{dv}{$d_v$}{VN degree}
\newacronym{dvs}{DVS}{dynamic voltage scaling}
\newacronym{dvb}{DVB}{digital video broadcast}
\newacronym{dl}{DL}{Deep Learning}
\newacronym{dnn}{DNN}{Deep Neural Network}
\newacronymf{dvbrcs}{DVB-RCS}{digital video broadcasting - return channel over satellite}
\newacronymf{dvbrct}{DVB-RCT}{digital video broadcasting - return channel terrestrial}
\newacronymf{dvbc2}{DVB-C2}{digital video broadcasting - cable - second generation \cite{eudigital}}
\newacronymf{dvbs2}{DVB-S2}{digital video broadcasting - satellite - second generation \cite{eudigital}}
\newacronymf{dvbs2x}{DVB-S2X}{digital video broadcasting - satellite - extension of the second generation \cite{eudigital}}
\newacronymf{dvbt2}{DVB-T2}{digital video broadcasting - terrestrial - second generation \cite{eudigital}}
\newacronym{dvfs}{DVFS}{dynamic voltage and frequency scaling}
\newacronym{drp}{DRP}{dithered relative prime}
\newacronym{dfg}{DFG}{Deutsche Forschungsgesellschaft}
\newacronym{dibco}{DIBCO}{Document Image Binarization Competition}
\newacronym{dsp}{DSP}{Digital Signal Processing}
\newacronym{dfki}{DFKI}{German Research Centre for Artificial Intelligence}
\newacronym{dpi}{DPI}{Dots Per Inch}
\newacronym{dnndk}{DNNDK}{Deep Neural Network Development Kit}
\newacronym{dpu}{DPU}{Deep Learning Processor Unit}

\newacronym{esf}{ESF}{extrinsic scaling factor}
\newacronym{edc}{EDC}{error detection and correction}
\newacronym{edge}{EDGE}{name of a communications standard}
\newacronym{eds}{EDS}{error detection sequential}
\newacronym{ecc}{ECC}{error correction code}
\newacronym{exit}{EXIT}{extrinsic information transfer}
\newacronym{ems}{EMS}{Extended Min-Sum}
\newacronym{ecn}{ECN}{Elementary Check Node}
\newacronym{ecg}{ECG}{Electrocardiogram}
\newacronym{ebr}{EBR}{Embedded Block RAM}

\newacronym{fhl}{FHL}{Forward Hidden Layer}
\newacronym{fsm}{FSM}{finite state machine}
\newacronym{fec}{FEC}{forward error correction}
\newacronym{fer}{FER}{Frame Error Rate}
\newacronym{fifo}{FIFO}{First In First Out memory}
\newacronym{fpga}{FPGA}{Field-Programmable Gate Array}
\newacronym{fsd}{FSD}{Fixed complexity sphere decoder}
\newacronym{fwbw}{FWBW}{Forward-Backward}
\newacronym{fcn}{FCN}{Fully Convolutional Network}
\newacronym{fpn}{FPN}{Feature Pyramid Network}
\newacronym{ff}{FF}{Flip-Flops}
\newacronym{fps}{FPS}{Frames Per Second}
\newacronym{fcnn}{FCNN}{Fully Connected Neural Network}
\newacronym{fc}{FC}{Fully Connected}

\newacronym{gsm}{GSM}{Global System for Mobile Communication; communication standard}
\newacronym{gpp}{GPP}{general-purpose processor}
\newacronym{gprs}{GPRS}{name of a communications standard}
\newacronym{gps}{GPS}{global positioning system}
\newacronym{geq}{GE}{gate equivalent}
\newacronym{gfq}{GF$(q)$}{Galois Field}
\newacronym{gpu}{GPU}{Graphics Processing Unit}
\newacronym{gpi}{GPI}{Global address space Programming Interface}
\newacronym{gops}{GOp/s}{Giga Operations per second}
\newacronym{gap}{GAP}{Global Average Pooling}

\newacronym{hspa}{HSPA}{high speed packet access}
\newacronym{harq}{HARQ}{hybrid automatic repeat request}
\newacronym{hw}{HW}{hardware}
\newacronym{hls}{HLS}{High-level Synthesis}
\newacronym{half}{HALF}{\underline{H}olistic \underline{A}uto machine \underline{L}earning for \underline{F}PGAs}
\newacronym{hdl}{HDL}{Hardware Description Language}

\newacronym{ic}{IC}{integrated circuit}
\newacronym{io}{I/O}{input / output}
\newacronym{ip}{IP}{Intellectual Property}
\newacronym{ier}{IER}{injected error rate}
\newacronym{itrs}{ITRS}{International Technology Roadmap for Semiconductors}
\newacronym{iot}{IoT}{Internet of Things}
\newacronym{ilsvrc}{ILSVRC}{ImageNet Large Scale Visual Recognition Challenge}

\newacronym{kde}{KDE}{Kernel Density Estimator}

\newacronym{lstm}{LSTM}{Long Short-Term Memory}
\newacronym{llr}{LLR}{Logarithmic Likelihood Ratio}
\newacronym{ldr}{LDR}{Logarithmic Density Ratio}
\newacronym{lte}{LTE}{long term evolution of the 3GPP standard}
\newacronym{ldpc}{LDPC}{Low-Density Parity-Check}
\newacronym{logmap}{Log-MAP}{MAP algorithm in the logarithmic domain}
\newacronym{lan}{LAN}{local area network }
\newacronym{lsb}{LSB}{least significant bit}
\newacronym{lfsr}{LFSR}{linear feedback shift register}
\newacronym{lord}{LORD}{layered orthogonal lattice detector}
\newacronym{lsd}{LSD}{List sphere decoder}
\newacronym{ldmc}{LDMC}{low-density MIMO check}
\newacronym{lut}{LUT}{LookUp Table}

\newacronym{mac}{MAC}{Multiply-Accumulate}
\newacronym{map}{MAP}{maximum a posteriori probability}
\newacronym{mlm}{Max-Log-MAP}{MAP algorithm in the logarithmic domain with simplified computation}
\newacronym{mimo}{MIMO}{multiple input multiple output}
\newacronym{mtbf}{MTBF}{mean time between failure}
\newacronym{msb}{MSB}{most significant bit}
\newglossaryentry{mpsoc}{name=MPSoC, description=multi-processor system-on-chip, first=multi-processor system-on-chip (MPSoC), firstplural=multi-processor systems-on-chip (MPSoC)}
\newacronym{mops}{MOPS}{million operations per second}
\newacronym{mmse}{MMSE}{minimum mean square error}
\newacronym{mcmc}{MCMC}{Markov Chain Monte Carlo}
\newacronym{mll}{ML}{maximum likelihood}
\newacronym{ml}{ML}{Machine Learning}
\newacronym{mdlstm}{MD-LSTM}{Multidimensional Long Short-Term Memory}
\newacronym{mlp}{MLP}{Multilayer Perceptron}
\newacronym{mdpi}{MDPI}{Multidisciplinary Digital Publishing Institute}

\newacronym{nns}{NNs}{Neural Networks}
\newacronym{nn}{NN}{Neural Network}
\newacronym{nbldpc}{NB-LDPC}{Non-Binary Low-Density Parity-Check}
\newacronym{nii}{NII}{next iteration initialization}
\newacronym{nsc}{NSC}{non-systematic and non-recursive convolutional}
\newacronym{noc}{NoC}{network on chip}
\newacronym{nom}{NOM}{nominal case, 1.2V, 25$^\circ$C}
\newacronym{nas}{NAS}{neural architecture search}
\newacronym{nlp}{NLP}{Natural Language Processing}

\newacronym{ocr}{OCR}{Optical Character Recognition}
\newacronym{otx}{OTX}{outer receiver}
\newacronym{ofdm}{OFDM}{orthogonal frequency division multiplexing}

\newacronym{pbb}{PBB}{Percentile-Based Binarization}
\newacronym{pl}{PL}{Programmable Logic}
\newacronym{ps}{PS}{Processing System}
\newacronym{psk}{PSK}{phase-shift keying}
\newacronym{psmap}{PSMAP}{parallel SMAP}
\newacronym{pr}{P\&R}{place and route}
\newacronym{ped}{PED}{partial Euclidean distance}
\newacronym{pic}{PIC}{parallel interference cancellation}
\newacronym{pn}{PN}{Permutation Node}
\newacronym{pim}{PIM}{Processing-In-Memory}
\newacronym{pcb}{PCB}{Printed Circuit Board}

\newacronym{qam}{QAM}{quadrature amplitude modulation}
\newacronym{qat}{QAT}{quantization aware training}
\newacronym{qbn}{QBN}{quantized batch norm}
\newacronym{qpsk}{QPSK}{quadrature phase shift keying}
\newacronym{qpp}{QPP}{quadratic permutation polynomial}
\newacronym{qos}{QoS}{Quality of Service}

\newacronym{rnn}{RNN}{Recurrent Neural Network}
\newacronym{rsc}{RSC}{recursive systematic convolutional}
\newglossaryentry{ram}{name=RAM,description=random access memory,first=random access memory (RAM),firstplural=random access memories (RAMs)}
\newglossaryentry{rom}{name=ROM,description=read only memory,first=read only memory (ROM),firstplural=read only memories (ROMs)}
\newacronym{rms}{RMS}{recognition, mining, and synthesis}
\newacronym{rs}{RS}{reed-solomon code}
\newacronym{rtl}{RTL}{register transfer level}
\newacronym{rts}{RTS}{repeated tree search}
\newacronym{rap}{RAP}{Resilience Articulation Point}
\newacronym{relu}{ReLU}{Rectified Linear Unit}
\newacronym{rl}{RL}{Reinforcement Learning}

\newacronym{snr}{SNR}{Signal-to-Noise Ratio}
\newglossaryentry{sram}{name=SRAM,description=static random access memory,first=static random access memory (SRAM),firstplural=static random access memories (SRAMs)}
\newacronym{smap}{SMAP}{serial MAP}
\newacronym{sdr}{SDR}{software-defined radio}
\newacronym{sdram}{SDRAM}{synchronous dynamic random access memory}
\newacronym{set}{SET}{single-event transient}
\newacronym{seu}{SEU}{single-event upset}
\newacronym{siso}{SISO}{soft-input soft-output}
\newacronym{siso2}{SISO}{single-input single-output}
\newacronym{soc}{SoC}{Systems-on-Chip}
\newacronym{sova}{SOVA}{soft-output Viterbi algorithm}
\newacronym{sw}{SW}{software}
\newacronym{sm}{SM}{spatial multiplexing}
\newacronym{stbc}{STBC}{space-time block code}
\newacronym{sttc}{STTC}{space-time trellis code}
\newacronym{sic}{SIC}{successive interference cancellation}
\newacronym{sts}{STS}{single tree search}
\newacronym{spp}{SPP}{Schwerpunktprogramm}
\newacronym{se}{SE}{Schnorr-Euchner}
\newacronym{synb}{SYN}{Syndrome-Based}
\newacronym{sgd}{SGD}{Stochastic Gradient Descent}
\newacronym{sdk}{SDK}{Software Development Kit}

\newacronym{tmr}{TMR}{triple modular redundancy}
\newacronym{tc}{TC}{turbo code}
\newacronym{ts}{TS}{Tuple search}
\newacronym{tpu}{TPU}{Tensor Processing Unit}

\newacronym{umts}{UMTS}{universal mobile telecommunications system}
\newacronym{uwb}{UWB}{ultra-wide band}
\newacronym{usb}{USB}{universal serial bus}
\newacronym{umic}{UMIC}{ultra high speed mobile information and communication}

\newacronym{vn}{VN}{Variable Node}
\newacronym{vfu}{VFU}{variable node functional unit}
\newacronym{vhdl}{VHDL}{very high speed integrated circuits hardware description language}
\newacronym{vliw}{VLIW}{very large instruction word}
\newacronym{vnb}{VNB}{variable node block (in \gls{LDPC} decoder architectures)}
\newacronym{vnp}{VNP}{variable node processors (in \gls{LDPC} decoder architectures)}
\newacronym{vblast}{V-BLAST}{Vertical Bell Labs layered space time}

\newacronym{wimax}{WiMAX}{worldwide interoperability for microwave access \cite{ieworldwide}}
\newacronym{wimedia}{WiMedia}{name of an alliance for standardization}
\newacronym{wifi}{WiFi}{trademark of Wi-Fi alliance}
\newacronym{wc}{WC}{worst case, 1.1V, 125$^\circ$C}

\newglossaryentry{xmap}{name=XMAP,description={name of a pipelined MAP decoder},first=XMAP,firstplural=XMAPs}
\newacronym{xor}{XOR}{exclusive OR function}

\newacronym{zf}{ZF}{zero-forcing}

\usepackage{mathtools}
\usepackage{amsmath}

\usepackage{threeparttable}

\usepackage{multirow}
\usepackage{rotating}
\usepackage{color}
\usepackage{makecell}

\usepackage{array,multirow,graphicx}
\usepackage{comment}

\usepackage{tikz}
\usepackage{pgf-pie}
\usepackage{varwidth}
\usetikzlibrary{positioning,shapes,fit,calc,patterns,patterns.meta}

\usepackage{wheelchart}
\usepackage {siunitx}
\usetikzlibrary{decorations.markings}

\definecolor{A0}{HTML}{A4DA90}
\definecolor{A1}{HTML}{81C269}
\definecolor{A2}{HTML}{74AA61}

\definecolor{B0}{HTML}{ECEB80}
\definecolor{B1}{HTML}{CFCD56}

\definecolor{C0}{HTML}{9D7AB3}
\definecolor{C1}{HTML}{845A9E}

\definecolor{D0}{HTML}{7E8BB4}
\definecolor{D1}{HTML}{5E6FA0}

\definecolor{E0}{HTML}{8D80B7}
\definecolor{E1}{HTML}{6F61A3}

\usepackage{todonotes}

\renewcommand{\baselinestretch}{0.93}

\usepackage{url}

\usepackage{pgfplots}
\pgfplotsset{compat=1.17}
\usepackage{pgfplotstable}
\usepackage[section]{placeins}

\usepackage{cite}

\usepackage{listings}
\usepackage{color}

\definecolor{dkgreen}{rgb}{0,0.6,0}
\definecolor{gray}{rgb}{0.5,0.5,0.5}
\definecolor{mauve}{rgb}{0.58,0,0.82}

\newcommand\cmxt{CardioMem® CM 100 XT}
\newcommand\half{proposed loop recorder platform}

\begin{document}


\title{Low-power, Energy-efficient, Cardiologist-level Atrial Fibrillation Detection for Wearable Devices}

\author{
  \IEEEauthorblockN{%
    Dominik Loroch\IEEEauthorrefmark{2}\IEEEauthorrefmark{1}\textsuperscript{\textsection}, 
    Johannes Feldmann \IEEEauthorrefmark{1}\textsuperscript{\textsection},
    Vladimir Rybalkin\IEEEauthorrefmark{1}\textsuperscript{\textsection},
    and Norbert Wehn\IEEEauthorrefmark{1}%
  }%
  \IEEEauthorblockA{\IEEEauthorrefmark{2} Fraunhofer ITWM, Kaiserslautern, Germany}
  \IEEEauthorblockA{\IEEEauthorrefmark{1} University of Kaiserslautern-Landau, Kaiserslautern, Germany}%
  \{dominik.loroch\}@itwm.fraunhofer.de, \{j.feldmann, rybalkin, wehn\}@rptu.de
}

\maketitle
\begingroup\renewcommand\thefootnote{\textsection}
\footnotetext{first three authors contributed equally}
\endgroup

\renewcommand{\baselinestretch}{1.0}
\begin{abstract}

Atrial fibrillation (AF) is a common arrhythmia and major risk factor for cardiovascular complications. While commercially available devices and supporting \gls{ai} algorithms exist for reliable detection of AF, the scaling of this technology to the amount of people who need this diagnosis is still a major challenge. 
This paper presents a novel wearable device, designed specifically for the early and reliable detection of AF. We present an FPGA-based patch-style wearable monitor with embedded deep learning-based AF detection. Operating with 3.8\,mW system power, which is 1-3 orders of magnitude lower than the state-of-the-art, the device enables continuous AF detection for over three weeks while achieving 95\% accuracy, surpassing cardiologist-level performance. 
A key innovation is the combination of energy-efficient hardware-software co-design and optimized power management through the application of hardware-aware neural architecture search. This advancement represents a significant step toward scalable, reliable, and sustainable AF monitoring.

\end{abstract}

\renewcommand{\baselinestretch}{0.935}

\section{Introduction}
\label{sec:intro}

\glsresetall

Cardiovascular diseases, including ischaemic heart disease and stroke, are the leading global causes of death \cite{who}. \gls{af}, the most common arrhythmia, is linked to serious outcomes such as stroke, heart failure, and mortality. Early, accurate detection is vital but challenging due to \gls{af}’s intermittent nature. Unlike standard short-term \gls{ecg}, \gls{aecg} devices provide continuous, long-term monitoring, improving the detection of infrequent yet critical arrhythmic events.

Most commercially available heart monitoring systems \cite{lifestar, ecardio, cardiopal, mcot, bodyguardian} are either one-piece devices with wired electrodes connected to a torso-mounted recorder, which are bulky and uncomfortable, or more recent two-piece systems comprising a chest-worn patch that wirelessly transmits \gls{ecg} data to a smartphone-like device. 
Some of these devices merely relay data, while others include proprietary algorithms for limited on-device analysis. However, in all cases, the wearable component primarily serves as a data transmitter. To our knowledge, we present the first patch-style wearable monitor with an embedded deep learning algorithm for near-sensor cardiologist-level \gls{af} detection. 
Ensuring comfort for long-term use demands compactness, which limits battery capacity and algorithm complexity, requiring power consumption in the milliwatt range. 
This paper presents a compact, low-power and energy-efficient solution enabling accurate, continuous \gls{af} detection for over three weeks.

\glspl{dnn} achieve cardiologist-level accuracy in arrhythmia detection \cite{hannun2019cardiologist}, but existing \gls{ecg} classification methods \cite{hannun2019cardiologist, kachuee2018ecg, jun2018ecg, saadatnejad2019lstm} either provide mediocre accuracy or require excessive computation, making them unsuitable for real-time use in energy- and power-constrained wearable devices. Software implementations on general-purpose processors are inefficient, necessitating custom hardware. 
\Glspl{fpga} offer benefits like flexibility, reconfigurability, and power efficiency.
Prior works implementing \gls{ecg} classification on an \gls{fpga} \cite{mangaraj2024fpga, karthikeyani2025framework, chandrasekaran2024fpga} use AMD/Xilinx \glspl{fpga}, which exhibit prohibitive power consumption for battery-operated wearables.
In contrast, we select the Lattice iCE40 UltraPlus for implementation.
Unlike high-end AMD/Xilinx \glspl{fpga} with abundant compute and memory resources, this device is highly resource-constrained, posing challenges for implementing cardiologist-level \gls{af} detection.
We explore how the unique characteristics of \glspl{fpga} can support this application and present the first \gls{fpga}-based \gls{aecg} patch-style device with embedded deep learning for \gls{af} detection.

Efficient \gls{dnn} deployment on a specific platform for a given cost criteria, requires navigating a vast design space, from network topology to hardware implementation, while accounting for complex cross-layer interdependencies. 
Manual optimization is infeasible, so this paper adopts a holistic cross-layer co-design methodology that jointly optimizes \glspl{dnn} and hardware. 
While some prior work focuses solely on maximizing prediction accuracy, others target efficient hardware implementations, overlooking how system-level constraints shape design. 
In contrast, our approach explicitly accounts for real-world conditions, including power and energy limits imposed by battery operation, data transmission overhead, on-board interfacing, ADC and external peripherals, thus enabling a truly practical and deployable solution.
An additional feature of this work is the handling of intrinsic and extrinsic noise sources during the hardware-aware neural architecture search, which is often overlooked in other work.



This study uses a novel dataset collected and annotated by medical professionals at Charité Hospital in Berlin. 
Compared to other, publicly available datasets for \gls{af}, our dataset takes special care of labeling noise patterns, such that model robustness towards noise and artifacts can be evaluated quantitatively, which enhances the reliability of the detection algorithms.

The contributions of this paper are the following:

\begin{itemize}
    \item We demonstrate a wearable patch-style monitor for near-sensor \gls{af} detection with above human predictive capabilities, reaching \textbf{95\% accuracy}.

    \item We design and implement an ultra low power \gls{fpga}-based platform, capable to execute \gls{dnn} within \textbf{3.8\,mW system power} consumption.

    \item We utilize hardware-aware \gls{nas} to find feasible \gls{dnn} solutions for our platform, utilizing only \textbf{14.6\,KByte of memory}.

    \item We create models robust against noise, with up to \textbf{98\% noise specificity}.
\end{itemize}

This paper contributes to advancing the state-of-the-art in intelligent device design, paving the way for more effective and accessible healthcare solutions.
\section{State-of-the-art}
\label{sec:sota}

Most of the work in the field of automatic \gls{ecg} analysis is dominated by \glspl{dnn}.
They have demonstrated superior performance in arrhythmia detection, outperforming human cardiologist, achieving accuracies of over 90\% \cite{hannun2019cardiologist}.
The most recent work uses multimodal models, fusing multiple representations of the 1D \gls{ecg} data into one input.
Also, 2D spectograms of the \gls{ecg} in combination with \glspl{cnn} are used to build classifiers for sequence or beat classification.
\cite{ahmad2021ecg} uses three Alexnets to preprocess several representations (GAF, MTF and RP image) and later combines the three latent representations in a fusion network to produce the final class label.
In contrast, \cite{xia2024multiscale} uses a multiscale dilated convolution network for \gls{af} detection on the raw \gls{ecg} data.
Another focus topic is to bring the algorithms onto embedded devices.
\cite{yazid2024atrial} uses a combination of local binary pattern feature extraction, which can be implemented efficiently using circular buffers, with SVMs and implements the algorithm on a STM32 microcontroller.
Using a combination of a VGG16 and CNN-BiLSTM, \cite{akter2024embedded} demonstrates a multimodal system running on a Raspberry Pi for the data processing, connected to a ESP8266 Microcontroller for data sampling, put together on a breadboard.

Lately, there are several publications targeting \glspl{fpga}.
\cite{mangaraj2024fpga} designs a hardware architecture for \gls{cnn}-based \gls{ecg} processing on a Xilinx Zynq FPGA, leveraging an array of processing elements to compute the convolution operation with intra-layer parallelism, but in a layer-by-layer manner.
In \cite{karthikeyani2025framework} the \gls{ecg} is processed by a graph \gls{cnn}, also using a processing element-based architecture for intra-layer parallelism on a Xilinx Artix \gls{fpga}.
There is an implementation of a vision transformer in combination with a capsule network by \cite{chandrasekaran2024fpga} on a Xilinx Zynq \gls{fpga}, using spectrograms of the \gls{ecg} signals as input and approximate computing for their MAC operations.

The models achieve high accuracies of up to 99\% across various detection tasks. However, this performance often comes at the cost of substantial model size and resource usage, making them impractical for ultra-low-power embedded devices. Consequently, \gls{fpga} implementations frequently rely on relatively large platforms such as the AMD/Xilinx Zynq. Model development typically prioritizes accuracy, with limited consideration for robustness to noise or constraints imposed by the target hardware. Hardware implementation is often treated as an afterthought, leading to a suboptimal platform choice. In this work, we tackle the problem from the other end and ask what a robust model needs to look like to operate effectively on a specific embedded device.

\section{Platform overview}
\label{sec:system_view}

As a reference, we use the commercially available \cmxt{}~\cite{cm100xt}, a medical-grade external loop recorder, widely employed by physicians to monitor patients with infrequent arrhythmic episodes.
It can operate for up to 14 days and records \gls{ecg} data when one of three events occurs: manual trigger by the patient, a time trigger set by the physician, or automatic detection of a cardiac anomaly via the Pan-Tompkins algorithm~\cite{pan_tompkins}, commonly used to detect QRS complexes in \glspl{ecg}.
Recordings are stored in flash memory and transmitted to the physician via Bluetooth using the patient’s mobile phone as a gateway, or via USB at the medical office.

A long runtime of the device increases the probability of detecting the \gls{af} sequences, which appear sporadically.
The targets for the new platform are 1) a sensitivity and specificity of at least 90\,\%, 2) a specificity under noise of at least 90\,\% and 3) a device runtime of at least 14 days.





\begin{figure}[!h]
    \centering
    \begin{tikzpicture}[thick, inner sep=0pt,
        scale=1, every node/.style={scale=0.8},
        rect_small/.style = {draw, rectangle, minimum height=8mm, minimum width=15mm},
        rect/.style = {draw, rectangle, minimum height=8mm, minimum width=20mm},
        square/.style = {draw, rectangle, minimum height=20mm, minimum width=20mm}
        ]

    \node[square] (FPGA) {FPGA};
    \node[rect] (EMMC) [below= 3mm of FPGA.south] {eMMC};
    \node[rect, right= 5mm of FPGA.south east, anchor=south west] (BUTTON) {Push button};
    \node[rect, right= 5mm of FPGA.north east, anchor=north west] (LED) {LED};
    \node[rect, above= 3mm of FPGA.north, anchor=south] (ADC) {ADC};
    \node[rect, align=center, right= 5mm of ADC.east, anchor=west] (ANA) {Analog\\Frontend};
    \node[rect, align=center, left= 5mm of FPGA.west] (BT) {Bluetooth\\SoC};
    \node[rect, above= 3mm of BT.north, anchor=south] (RTC) {RTC};
    \node[rect, below= 3mm of BT.south, anchor=north] (USB) {USB};
    \node[rect, align=center, left= 5mm of BT.west, anchor=east] (Lipo) {LIPO\\Charger};

    \draw[-] (FPGA) -- (BUTTON);
    \draw[-] (FPGA) -- (LED);
    \draw[-, line width=2pt] (FPGA) -- (ADC);
    \draw[-] (ADC) -- (ANA);
    \draw[-, line width=2pt] (FPGA) -- (EMMC);
    \draw[-, line width=2pt] (FPGA) -- (BT);
    \draw[-, line width=2pt] (BT) -- (RTC);
    \draw[-, line width=2pt] (BT) -- (USB);
    \draw[-, line width=2pt] (BT) -- (Lipo);
    
    \end{tikzpicture}
    
    \caption{Overview of the ECG loop recorder platform}
    \label{fig:HWArch}
    \vspace{-2pt}
\end{figure}
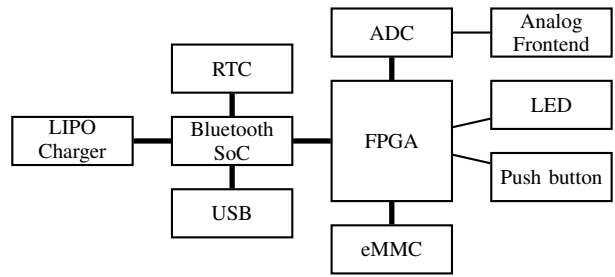

An abstract architectural overview of the proposed loop recorder platform is shown in \cref{fig:HWArch}.
The platform comprises two programmable devices: a Lattice iCE40 Ultra-Plus FPGA (iCE40UP5K) and a Silicon Labs BGM240 Series Bluetooth module (BGM240PA22VNA).
Leveraging hardware-software partitioning, event-based processing is handled by the BGM240, while continuous data processing is performed by the FPGA. 
The ultra-low-power FPGA continuously gathers ECG data sampled by the ADC, storing it in internal SRAM and analyzing it using a DNN in $120\,s$ windows.
After each analysis, the neural network determines whether a \gls{af} is present.
If detected, the FPGA signals the event to the Bluetooth module via an interrupt and flashes an LED to alert the patient.
\Gls{ecg} data remains in SRAM until transferred to the Bluetooth module.
If no \gls{af} is detected or the transfer completes, the FPGA clears memory and resumes data acquisition.
The Bluetooth module, integrating an ARM Cortex M33 and an ARM Cortex M0+ as radio controller, handles wireless communication and patient interaction.
On receiving an interrupt, it transfers data from the FPGA via SPI and appends a timestamp using the real-time clock (RTC).
Data is transmitted to the hospital via Bluetooth, using the patient’s smartphone as a gateway.
If the phone is unavailable, data is stored in an eMMC device. Since the BGM240 lacks an eMMC controller, one is implemented on the \gls{fpga}, allowing data transfer back to the \gls{fpga} via a second SPI.
Once in the eMMC, data can be sent when connectivity is restored or read via USB using an FTDI USB-to-UART interface.
A push button on the front of the loop recorder allows the patient to initiate ECG data transfer manually, if feeling unwell, regardless of the neural network’s decision.

Unlike the \cmxt{}, which is powered by a replaceable Renata CR2477N 3V Lithium battery, our proposed platform uses a Lithium polymer battery (LIPO) with similar energy capacity.
This change was necessary due to power supply limitations for the eMMC.

\begin{figure}[!h]
\centering
\includegraphics[width=1.0\columnwidth]{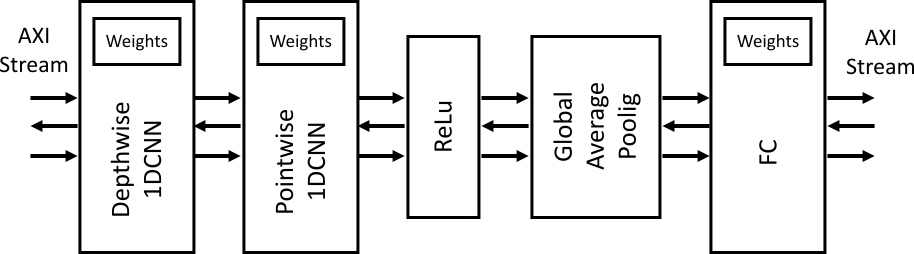}
\caption{The sample hardware architecture.}
\label{fig:hw_architecture}
\vspace{-12pt}
\end{figure}

The selected \gls{fpga} has three types of on-chip memory resources, namely distributed RAM, \gls{ebr}, and SPRAM primitives. The use of external memory was omitted to reduce power consumption. 
As all SPRAM is allocated for loop recording, only distributed RAM and \gls{ebr} remain for storing the \gls{dnn} model parameters and feature maps. 
This constraint necessitates a streaming architecture, where data flows directly between layers, unlike traditional layer-by-layer processing that relies on large intermediate buffers. 
The \gls{fpga} implementation leverages a customizable \gls{hls}-based hardware library comprising modular C++ templates for various \gls{dnn} layers, based on the architecture introduced in \cite{ney2021half}, with multiple enhancements to improve resource efficiency as explained in \cref{sec:results}.
Designed for low power and ultra-low latency, the architecture stores all weights and intermediate data in on-chip memory to minimize energy use and delay. 
Fully pipelined and dataflow-driven, it enables concurrent layer execution and immediate computation upon input availability. 
Streaming interfaces and modular design promote rapid development, debugging, and integration, with individual layer modules interconnected via on-chip data streams in a unified top-level module depicted in \cref{fig:hw_architecture}.

\section{Methodology}
\label{sec:methodology}
\begin{figure*}[!t]
    \centering
    \begin{tikzpicture}[scale=0.85] 
    \begin{axis}[
        axis line style={draw=none}, 
        tick style={draw=none},
        ybar,
        bar width=10pt,
        width=\textwidth,
        height=0.35\textwidth,
        enlargelimits=0.15,
        legend style={at={(0.98,0.98)}, anchor=north east},
        ylabel={Resource Utilization (\%)},
        symbolic x coords={Baseline,Opt. 1,Opt. 2,Opt. 3,Opt. 4,Opt. 5,Final},
        xtick=data,
        x tick label style={yshift=20pt},
        ymin=0, ymax=300,
        ymajorgrids=true,
        nodes near coords,
        nodes near coords align={vertical},
        legend cell align={left}
    ]
    
    \addplot+[style={blue,fill=blue!50}] 
    coordinates {(Baseline, 114) (Opt. 1, 86) (Opt. 2, 78) (Opt. 3, 75) (Opt. 4, 51) (Opt. 5, 52) (Final, 35)};
    
    \addplot+[style={red,fill=red!70}] 
    coordinates {(Baseline, 305) (Opt. 1, 260) (Opt. 2, 248) (Opt. 3, 228) (Opt. 4, 123) (Opt. 5, 106) (Final, 72)};
    
    \addplot+[style={green!70!black,fill=green!40}] 
    coordinates {(Baseline, 100) (Opt. 1, 100) (Opt. 2, 100) (Opt. 3, 100) (Opt. 4, 100) (Opt. 5, 87) (Final, 62)};
    
    \addplot+[style={yellow!80!black,fill=yellow!60}] 
    coordinates {(Baseline, 113) (Opt. 1, 80) (Opt. 2, 113) (Opt. 3, 106) (Opt. 4, 90) (Opt. 5, 106) (Final, 43)};
    
    \legend{FF,LUT,DSP,EBR}
    \end{axis}
    \end{tikzpicture}
    \caption{The resource utilization of various configurations on Lattice iCE40UP5K.}
    \label{fig:resource_utilization}
    \vspace{-12pt}
\end{figure*}
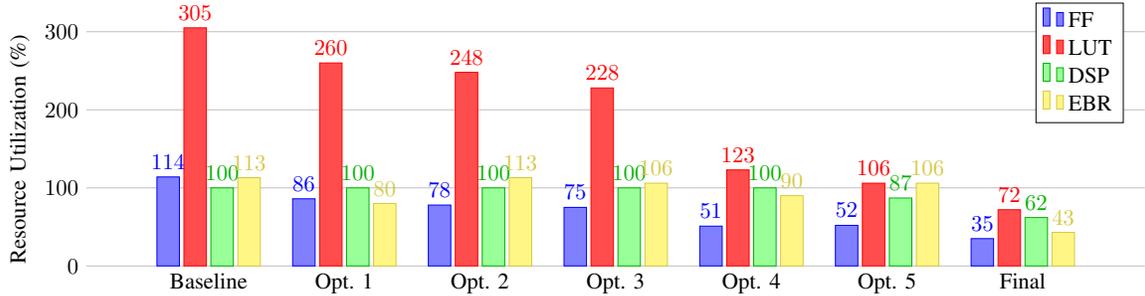
The goal is to find a \gls{dnn} topology, that satisfies: 1) sensitivity and specificity requirements for the application, 2) is robust against noisy input data and 3) runs given the constraints from the \gls{fpga} platform and hardware architecture.
Sensitivity and specificity are clearly defined as $\frac{TP}{TP+FN}$ and $\frac{TN}{TN+FP}$. These two metrics describe the model sufficiently in presence of imbalanced test data, in contrast to accuracy and precision.

Robustness is measured as the specificity when the input consists only of noisy data.
Sensitivity is not considered in this context, because the noise makes a clear identification of a positive feature difficult or impossible.
We distinguish two sources of noise: intrinsic and extrinsic noise.
Intrinsic noise originates from the processing of the signal, in our case primarily quantization noise from the fixed-point arithmetic used on the \gls{fpga}.
The noise level is controlled directly by \gls{dnn} design, i.e. by the bitwidth and precision of the fixed-point format. 
Extrinsic noise originates from sources outside of the device, like bad leads or triboelectricity, which is specific to \gls{ecg} data.

Finally, the model needs to be implemented on the target \gls{fpga}.
The entire task can be posed as a multi-objective optimization problem.
We use \gls{nas} to find a suitable topology for the \gls{fpga} platform that meets all requirements.
The basis of the implementation in this work is a genetic, evolutionary \gls{nas} algorithm introduced in \cite{ney2021half}, that is especially well suited for hardware-aware multi-objective optimization.
For details on the objective function and selection process we refer to that paper.
In this paper, we extend the \gls{nas} with specific targets and constraints for the selected \gls{fpga} platform, hardware architecture and the noise handling.

Given the resource constraints of the platform, our search space consists of very flat networks.
The primary building block are 1D depthwise separable convolution layers, which splits a single regular convolution into two steps with
\begin{equation}
\begin{aligned}
    \mathrm{DSConv}(X) &= ( X \ast W_d ) \ast W_p,
\end{aligned}
\end{equation}
where $X\in \mathbb{R}^{H\times C_\mathrm{in}}$ is the input, $W_d\in \mathbb{R}^{K\times C_\mathrm{in}}$ the depthwise kernel weights, $W_p\in \mathbb{R}^{C_\mathrm{in} \times C_\mathrm{out}}$ the pointwise kernel weights and "$\ast$" denotes the 1D convolution operation.

Depthwise separable convolutions reduce the computational cost from $H\times K \times C_\mathrm{in} \times C_\mathrm{out}$ to $H\times C_{in}(K+C_\mathrm{out})$, where $H$ is the length of the input sequence and $K$ is the kernel size. 
This difference can be quite substantial if $K$ or $C_\mathrm{out}$ is large, which is the case for our topologies.
Additionally, there is the stride $S$, which reduces the output size to $H/S$ by skipping output elements.

We train our networks using \gls{qat}. 
The \gls{fpga} platform supports fixed point quantization of arbitrary word width $w$ and precision $p$.
We define a quantization operator as
\begin{equation}
\begin{aligned}
    &Q_{w,p}(x) = \\ &\mathrm{clip}\left(\mathrm{sign}(x) \frac{\lfloor |x|*2^p+0.5)\rfloor}{2^p}, -2^{w-p-1}, 2^{w-p-1}-1\right).
\end{aligned}
\end{equation}
We use straight-through estimators for the backpropagation of the non-differentiable $Q_{w,p}()$ operator.
The quantized convolution operator for training the networks with \gls{qat} is 
\begin{equation}
    \mathrm{QDSConv}(X) = Q_\mathrm{a}\left(Q_\mathrm{a}( Q_\mathrm{a}(X) \ast Q_\mathrm{w}(W_d)) \ast Q_\mathrm{w}(W_p)\right),
\end{equation}
where $Q_\mathrm{w}$ is a short form for $Q_{w_\mathrm{w},p_\mathrm{w}}$ for the weight quantization and its parameters, respectively for the activations.

Every convolution layer is followed by a quantized batch normalization layer
\begin{equation}
    \mathrm{QBatchNorm}(X) = \frac{Q_\mathrm{a}(X)-Q_\mathrm{w}(\mu_X)}{\sqrt{Q_\mathrm{w}(\sigma_X)^2+\epsilon}} Q_\mathrm{w}(\gamma) + Q_\mathrm{w}(\beta),
\end{equation}
with $\mu_X$ the moving average, $\sigma_X^2$ the moving variance, $\gamma$ a trainable scaling factor and $\beta$ a bias value.
We found that batch normalization is very substantial to the training of the networks, but quantization in the batchnorm is necessary for batchnorm folding after training, which additionally saves resources on the \gls{fpga}.
All of our networks end with a \gls{gap} layer, followed by a simple \gls{fc} layer that produces a single logit for binary classification.
    
The search space for the convolution layers is defined by $K\in$ \{1, 2, 4, 8, 16, 32, 64, 128, 256\}, $C\in$ \{4, 8, 16, 32, 64, 128, 256, 512, 1024\}, $S\in$ \{1, 2, 4, 8, 16, 32, 64\} and $Q_\mathrm{w,a}\in$ \{[32,16], [24,16], [16,10], [16,8], [16,12],[12,6], [12,8]\}.
Note that the parameters for $K$, $C$ and $S$ are constraint to be powers of 2 to simplify hardware implementation of the loop logic.
    

Additionally, we define optimization targets and respective constraints for the \gls{nas} to force hardware-awareness in the topologies:
1) The number of model parameters is $\le10^6$.
2) The number of convolution layers is minimized and $\le5$.
3) Sensitivity and specificity are maximized with a lower constraint of $0.7$.
We relax the lower bound, so the \gls{nas} can explore the search space more flexibly, which leads to better designs in the end.
4) Noisy specificity, also with a lower constraint of $0.7$.
5) Quantization word width and precision, for both weights and activations, are minimized in bit size, reducing resource requirements on the \gls{fpga}.
6) Output size $H\times C_\mathrm{out}$ of every layer of the network is minimized, reducing resource utilization.

\section{Results}
\label{sec:results}

\paragraph{Co-design}
This section demonstrates the accumulated effect of \gls{dnn} design space exploration and hardware architecture enhancements on the \gls{fpga} resource utilization.

\textbf{Baseline:} The initial topology consists of a 5 layer network.
Hardware implementation of this design exceeds the available \gls{fpga} resources (see \cref{fig:resource_utilization}), prompting the need for architectural and topological optimizations. Resource analysis reveals that depthwise separable convolutions account for approximately 80\,\% of \gls{lut}  usage, identifying them as the primary bottleneck.

\textbf{Opt. 1:} In the baseline hardware architecture, convolution layers comprise of two units. One is the \gls{mac} unit and the other is a line buffer that prepares the input feature map for the \gls{mac} unit to comply with the computation pattern. For pointwise convolutions, the line buffer was removed due to the relatively simple reading pattern of the pointwise convolution, reducing resource usage. For depthwise convolutions, constraining strides $\leq$ kernel size, minimized control logic and memory requirements, yielding a 15\% improvement in resource utilization.

\textbf{Opt. 2:} To optimize \gls{mac} resource usage, \gls{nas} was constrained to use kernel sizes and number of input and output channels as powers of two, enhancing \gls{lut} efficiency for loop logic, however, at the cost of larger kernels (\{5,3,8,3,4\} $\rightarrow$ \{8,4,8,4,8\}) and slightly increased \gls{ebr} usage. Additionally, fusing the ReLU with the preceding pointwise layer eliminated module overhead, yielding a further 5\,\% \gls{lut} reduction.

\textbf{Opt. 3:}  The \gls{gap} and \gls{fc} layers accounted for 12\,\% of \gls{lut} usage. By fusing these layers, \gls{lut} utilization was reduced by an additional 8\,\% compared to the previous optimization.

\textbf{Opt. 4:} Quantization was applied to weights, biases, activations, and internal \gls{mac} results to further reduce resource usage. Variable ranges were profiled using training and validation datasets for post-training quantization to determine the required integer bitwidth (int).
Total bitwidths (int+precision) were then set as: weights/biases (int+6), \gls{mac} results (int+12), and activations (int+8). This reduced resource utilization by 46\,\%, though still insufficient for \gls{fpga} placement.

\textbf{Opt. 5:} In the following, we selected a 3-layer model. The compact architecture with no accuracy loss was found by increasing kernel sizes (\{8,4,8,4,8\} $\rightarrow$ \{16,64,128\}), resulting in marginal \gls{lut} savings but increased \gls{ebr} utilization due to the expanded receptive fields.

\textbf{Final:} Aware of the resource cost of large kernels, we added kernel size limits to previous constraints. The final model is a 2-layer model, reducing \gls{lut} usage by 32\,\% over the prior design and leaving sufficient resources for full system integration.

\paragraph{NAS}
In order to evaluate the noisy specificity, we require a dataset that contains samples of \gls{af} and noise, both labeled accordingly.
A new dataset was created, using the GE SEER 1000 device, from 30 probands wearing the device for 1 to 3 weeks, resulting in approx. 98K samples of 120s length.
The data was annotated by medical staff, where special focus was put on also labeling noisy sequences.
The dataset is split in 70\,\% for training, 15\,\% for validation and 15\,\% for final testing.
During the \gls{nas}, the networks were all trained using stochastic gradient descent with momentum, Nesterov acceleration and gradient clipping for 30 epochs.
We used a learning rate schedule starting from $0.01$ and dividing by 10 at epoch 15 and 25.
Inspired by \cite{rahman2023systematic}, we augmented the data by adding low amplitude random noise (25\,Hz-100\,Hz), a zero-line shift ($\pm 10\%$), signal amplitude scaling ($\pm 20\,\%$) and frequency shift ($\pm 10\%$).
We ran the \gls{nas} for 190 generations, producing 8 offsprings per generation, starting from randomly generated topologies in generation 1.

\begin{figure}[!htbp]
    \centering
    \vspace{-4pt}
    \includegraphics[width=.8\columnwidth]{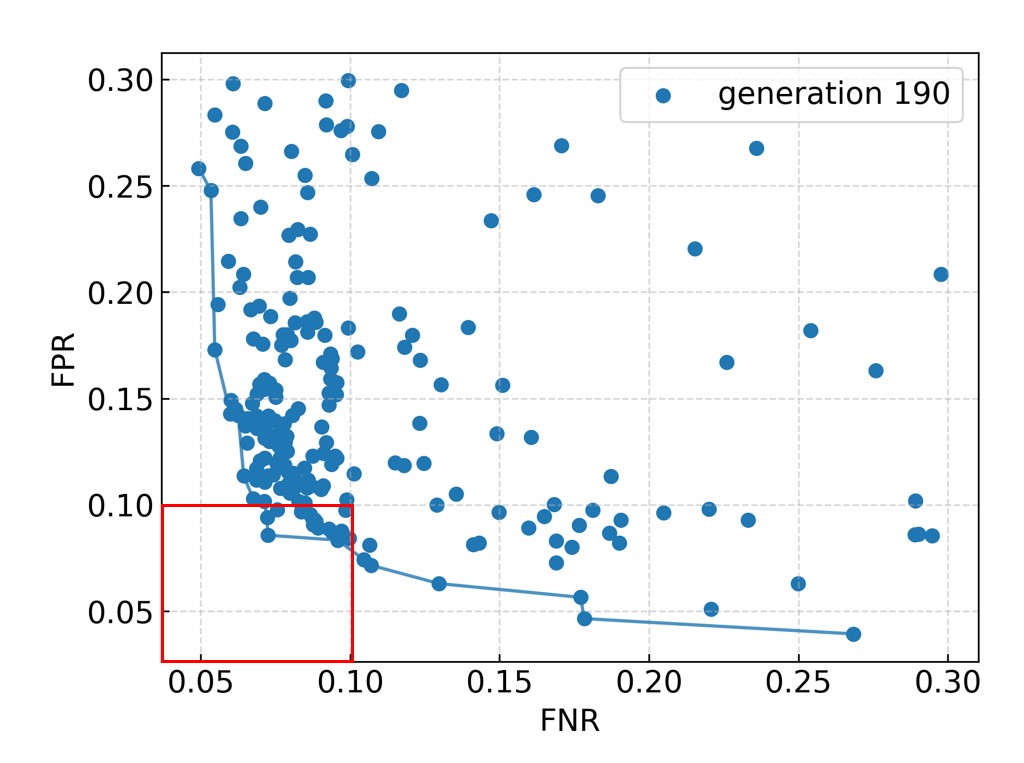}
    \caption{ False negative rate FNR ($1-$sensitivity) vs. false positive rate FPR ($1-$specificity) for the Pareto set of the last generation of the \gls{nas}.}
    \label{fig:pareto_190}
    \vspace{-2pt}
\end{figure}
The last generation contains 220 Pareto optimal individuals shown in \cref{fig:pareto_190}.
We observed the first individuals with 90\,\% sensitivity and specificity begin to show up around generation 70 and improving marginally until the end.

\begin{table}[!tbhp]
    \centering
    \caption{Performance measurements for the models with deviation in paranthesis.}
    \label{tab:model_perf}
    \begin{tabularx}{\columnwidth}{*{5}{>{\centering\arraybackslash}X}}
    \toprule
        Model & Sensitivity & Specificity & Noise spec. \\
    \midrule
        1:  & 0.899 & 0.934 & 0.946 \\
        25856 params& (0.026) & (0.025) & (0.007) \\
                      &&& \\
        2:  & 0.903 & 0.968 & 0.970 \\
        4644 params              & (0.029) & (0.004) & (0.010) \\
                      &&& \\
        3:  & \textbf{0.938} & \textbf{0.981} & \textbf{0.985} \\
        7328 params  & (0.007) & (0.003) & (0.006) \\
    \bottomrule
    \end{tabularx}
    \vspace{-4pt}
\end{table}
We perform an analysis on three well performing topologies selected from the Pareto set and compare their predictive performance in \cref{tab:model_perf}.
We estimate the values by training each model five times and reporting the empirical mean and standard deviation of the measured values.
Remarkably, model 3 outperforms all the other models by a significant margin, with a very low deviation.
Model 1 is the worst, despite being the biggest model by a large margin.
All found models are resilient against noise, which stems from the consideration of quantization noise and the noise specificity during the search.

\begin{table}[!htb]
    \centering
    \caption{Comparison of the model performance with deviation across different datasets.}
    \label{tab:model_datasets}
    \begin{tabularx}{\columnwidth}{*{5}{>{\centering\arraybackslash}X}}
    \toprule
        Model & Our dataset & TIM-HF2  & MIT BIH  & PTB-XL \\
              &             & \cite{koehler2018efficacy} & \cite{moody2001impact} & \cite{wagner2020ptb} \\
    \midrule
        1:  & 0.907(0.011) & 0.766(0.025) & 0.883(0.049) & input too\\
                      & 0.927(0.019) & 0.564(0.072) & 0.924(0.017) & small \\
                      &&&&\\
        2:  & 0.910(0.024) & 0.854(0.036) & \textbf{0.975(0.026)} & 0.863(0.016)\\
                      & 0.956(0.027) & 0.799(0.018) & 0.906(0.020) & 0.811(0.046)\\
                      &&&&\\
        3:  & \textbf{0.937(0.006)} & \textbf{0.946(0.010)} & 0.958(0.001) & \textbf{0.931(0.010)}\\
                      & \textbf{0.978(0.003)} & \textbf{0.922(0.030)} & \textbf{0.958(0.005)} & \textbf{0.937(0.004)}\\
    \bottomrule
    \end{tabularx}
    \vspace{-4pt}
\end{table}
Possibly, the best performing model could be topologically overfitted to the dataset we run the \gls{nas} on. Table \ref{tab:model_datasets} shows the evaluation of the sensitivity and specificity of the models on different datasets.
The training parameters are the same.
We use only the first two channels of each dataset, if it has more than 2 channels, without observing loss in model performance.
Model 3 achieves comparable accuracy across all datasets without much change in performance.
Other models struggle more to adapt to a different dataset.
Although the \gls{nas} found indeed a model which is robust not only for the dataset it has been initially designed for, but also transfers onto other datasets, recorded with other devices and sampling frequencies, it is not a property that automatically emerges in all of the models.
An additional optimization criterion should be introduced to find transferable models, e.g. performance across multiple test datasets, if this property is required.

\begin{figure}[!hb]
    \centering
    \vspace{-12pt}
    \begin{tikzpicture}[node distance={20mm and 20mm}, thick, inner sep=2pt,
        label/.style = {},
        ]

     \node[anchor=south west,inner sep=0] at (0,0) {
        \includegraphics[trim={0 10cm 0 8cm},clip,width=1.0\columnwidth]{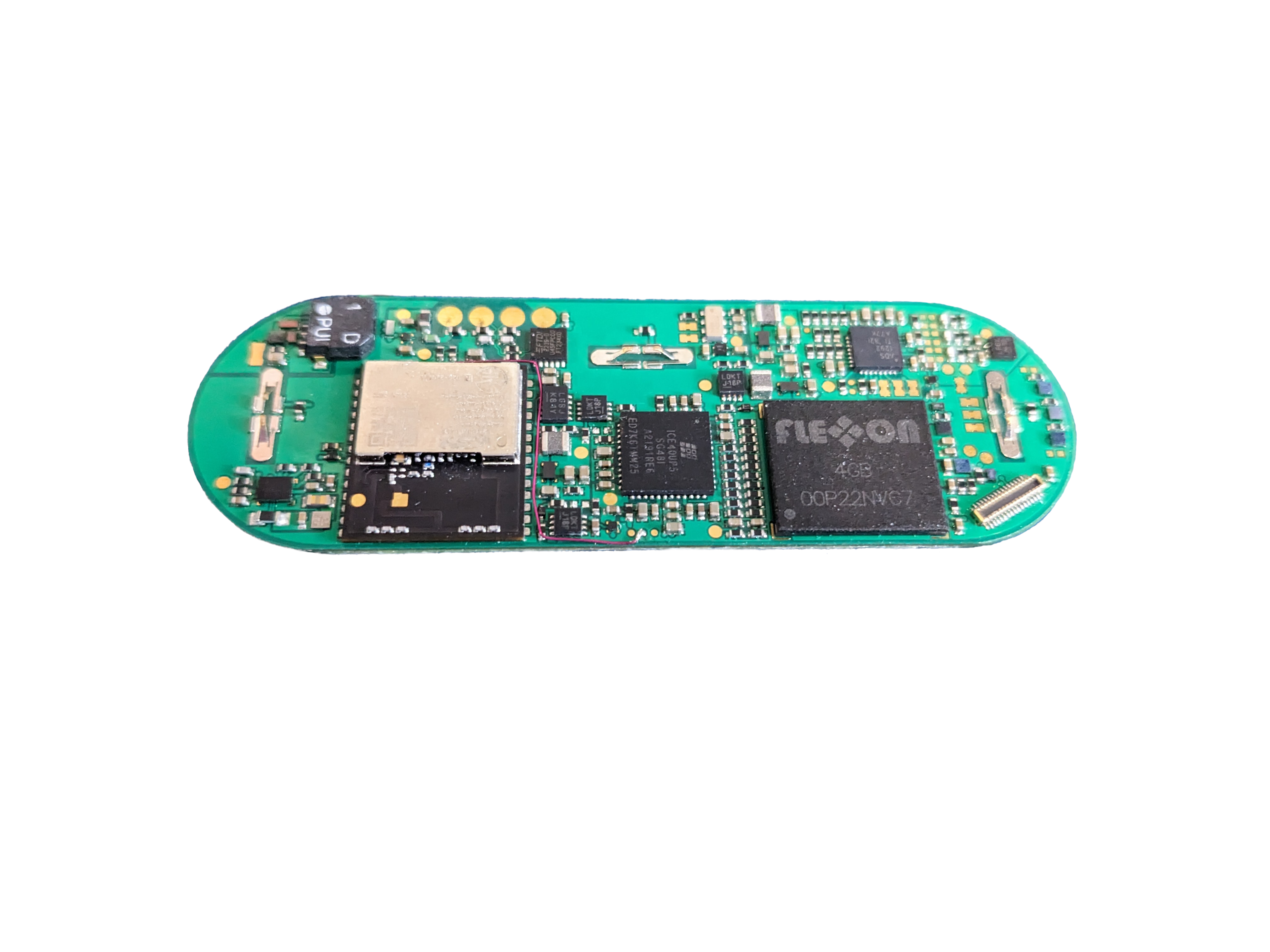}
     };

    \draw[line width=0.3mm, |<->|, gray] (1.3, 2.2) -- (7.3, 2.2) node [near start, above] (TextNode1) {7.0\,cm};
    \draw[line width=0.3mm, |<->|, gray] (1.1, 2) -- (1.1, 0.2) node [midway, above, rotate=90] (TextNode2) {2.3\,cm};

    \node[label] (FPGA) at (4.5,2.5) {FPGA};
    \node[label] (MCU) at (2.5,-0.5) {BGM240};
    \node[label] (ADC) at (6.5,2.5) {ADC};
    \node[label] (EMMC) at (6.5,-0.5) {eMMC};
    
    \draw[line width=0.5mm, red] (FPGA) -- (4.6,1);
    \draw[line width=0.5mm, red] (MCU) -- (3,1);
    \draw[line width=0.5mm, red] (ADC) -- (6.1,1.6);
    \draw[line width=0.5mm, red] (EMMC) -- (6,0.8);

    \end{tikzpicture}
    \caption{Manufactured \gls{pcb} of the \half{}. Weight incl. Battery: 10\,g.}
    \label{fig:loop_recorder}
    \vspace{-6pt}
\end{figure}

\paragraph{Hardware}
We compare the proposed loop recorder platform, shown in \cref{fig:loop_recorder}, with the external ECG loop recorder \cmxt{}\cite{cm100xt}.
Power measurements were conducted using a Rohde\&Schwarz NGU201 source measurement unit.
The proposed platform, powered by a 3.7\,V LIPO battery, was tested at its nominal voltage.
In contrast, the CardioMem® CM 100 XT uses a 3\,V Renata CR2477N Lithium battery with similar energy capacity.
The proposed platform draws 1.03\,mA to acquire and analyze ECG data without Bluetooth transmission or eMMC storage, resulting in 3.81 mW power consumption — 26.3\,\% lower than the 5.17\,mW drawn by the CardioMem® CM 100 XT. Consequently, the proposed platform can continuously analyze ECG data for over 22 days, 57\,\% longer than the CardioMem® device. During an event, it consumes an additional 418\,mJ in 46\,s, equivalent to 109.68\,s of loop recording. So every event reduces the total runtime by under two minutes.



\begin{figure}[htbp]
    \centering

    
    \includegraphics[width=0.9\columnwidth]{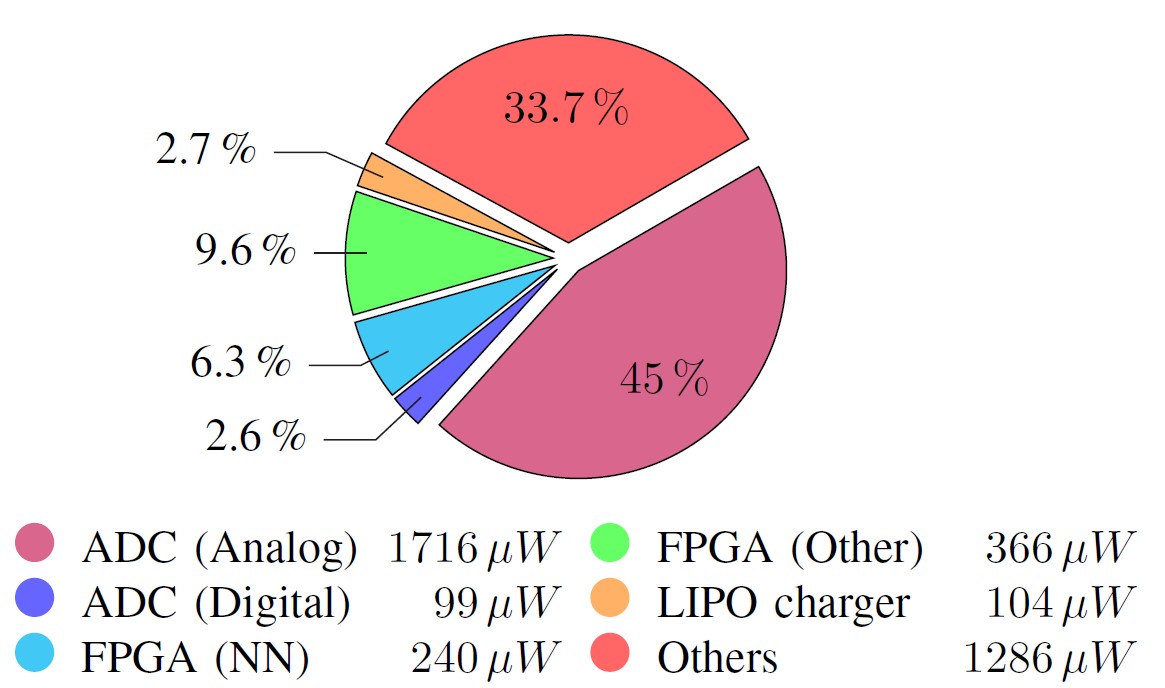}
    \vspace{4pt}
    \caption{Distribution of power consumption of the \half{} while loop recording.}
    \label{fig:power_dist}
    \vspace{-12pt}
\end{figure}

\cref{fig:power_dist} shows the power consumption distribution of the \half{} during loop recording. Nearly half the power is consumed by the analog and digital domains of the ADC for continuous ECG sampling and transfer.
The FPGA accounts for $15.9\,\%$ of total power — $6.3\,\%$ for the neural network and $9.6\,\%$ for loop recorder functions, configuration/status registers, and clock management.
As the platform is powered by a LiPo battery, a charger is required, consuming $2.7\,\%$ of power to continuously monitor battery voltage and current.
The \textit{Others} category includes the BGM240 Bluetooth module, push-button controller, power monitors, and a DC/DC converter, which are not discussed in this paper.


\begin{table}[htbp]
    \centering
    \caption{Comparison to the state of the art. se: sensitivity, sp: specificity, acc: accuracy.}
    \label{tab:sota}
    \setlength{\tabcolsep}{3pt}
    \begin{tabularx}{\columnwidth}{Xccccc}
        \toprule
        Work & Size & \multicolumn{3}{c}{Performance} & Power \\
             & [params] & se & sp & acc & [mW] \\
        \midrule 
        \cite{xia2024multiscale} CNN & 32M & \textbf{99.5}\% & \textbf{99.6}\% & - & - \\
        \cite{ahmad2021ecg} CNN & 60M & 98.0\% & - & 99.2\% & - \\
        \cite{mangaraj2024fpga}  CNN/FPGA & 15.4K & - & - & 98.6\% & 4170\\
        \cite{akter2024embedded}  BiLSTM/RaspPi & $>$165K & 92.0\% & - & 92.7\% & $>$3000\\
        \cite{karthikeyani2025framework}  GCNN/FPGA & - & 96.3\% & 98.8\% & - & 1570\\
        \cite{chandrasekaran2024fpga}  ViT/FPGA & - & 97.5\% & - & - & 1020\\
        \cite{yazid2024atrial}  SVM/MCU & 114\,KByte & \textbf{99.5}\% & 99.2\% & - & 89\\
        \midrule
        \textbf{Ours} NAS-CNN/FPGA &  \textbf{7.3K} & 95.8\% & 95.8\% & - & \textbf{3.81} \\
                                     & (14.6\,KByte) &&&&\\
        \bottomrule
    \end{tabularx}
    \vspace{-2pt}
\end{table}

A comparison with state-of-the-art models and implementations for \gls{ecg} analysis is given in Table \ref{tab:sota}.
Our model performs with comparable performance to other work, but with 1-4 orders of magnitude smaller model size and 1-3 orders of magnitude smaller system power consumption.
We attribute this success to the use of hardware-aware \gls{nas}, which automatized the model design given the ultra-low-power hardware platform and its constraint resources.
Additionally, we have shown that our model is robust to noise and achieves consistent performance across multiple datasets.


\section{Conclusion}
\label{sec:conclusion}


We demonstrated an ultra low power wearable device for precise and reliable \gls{af} detection on the edge.
We achieve state-of-the-art accuracy at a power level that is at least one order of magnitude below most other work in this field.
A special focus on robustness towards noisy data demonstrates that our solution is not only compact, but also consistent.
We achieved these goals by utilizing the most state-of-the-art hardware-software co-design workflow, showing the potential of this methodology.
Furthermore, our methodology is applicable to many other domains, such as predictive maintenance for industry 4.0.

\section*{acknowledgment}
We would like to thank GETEMED Medizin- und Informationstechnik AG for their support and valuable feedback in building the demonstrator platform with us.
We thank Charite DHZC for giving us access to previous studies, as well as planning and conducting a new study with us to enable the findings of this paper.
This study was conducted under the BMBF grant 16ME0387K and partially supported by the Carl Zeiss
Stiftung, Germany, under the Sustainable Embedded AI project (P2021-02-009).

\FloatBarrier

\bibliographystyle{ieee/IEEEtran} 
\bibliography{ieee/IEEEexample}


\end{document}